# Commensurate antiferromagnetic excitations as a signature of the pseudogap in the tetragonal high-$T_c$ cuprate $HgBa_2CuO_{4+\delta}$

M. K. Chan[1,2], C. J. Dorow[1,a], L. Mangin-Thro[3], Y. Tang[1], Y. Ge[1,b], M. J. Veit[1,c], G. Yu[1], X. Zhao[1,4], A. D. Christianson[5], J. T. Park[6], Y. Sidis[3], P. Steffens[7], D. L. Abernathy[5], P. Bourges[3], and M. Greven[1]

[1] School of Physics and Astronomy, University of Minnesota, Minneapolis, MN 55455, USA

[2] Mail Stop E536, Pulsed Field Facility, National High Magnetic Field Laboratory, Los Alamos National Laboratory, Los Alamos, NM 87545, USA

[3] Laboratoire Léon Brillouin, LLB/IRAMIS, UMR12, CEA-CNRS, CEA-Saclay, 91191 Gif sur Yvette, France

[4] State Key Lab of Inorganic Synthesis and Preparative Chemistry, College of Chemistry, Jilin University, Changchun 130012, China

[5] Oak Ridge National Laboratory, Oak Ridge, TN 37831, USA

[6] Forschungsneutronenquelle Heinz Maier-Leibnitz, 85747 Garching, Germany

[7] Institut Laue Langevin, 38042 Grenoble CEDEX 9, France

[a] Present address: Department of Physics, University of California, San Diego, 9500 Gilman Drive La Jolla, CA 92093, USA

[b] Present address: Department of Physics, Penn State University, University Park, PA 16802, USA

[c] Present address: Department of Applied Physics, Stanford University, Stanford, CA 94305, USA

**Antiferromagnetic correlations have been argued to be the cause of the *d*-wave superconductivity and the pseudogap phenomena exhibited by the cuprates. Although the antiferromagnetic response in the pseudogap state has been reported for a number of compounds, there exists no information for structurally simple $HgBa_2CuO_{4+\delta}$. Here we report neutron scattering results for $HgBa_2CuO_{4+\delta}$ (superconducting transition temperature $T_c \approx 71$ K, pseudogap temperature $T^* \approx 305$ K) that demonstrate the absence of the two most prominent features of the magnetic excitation spectrum of the cuprates: the X-shaped 'hourglass' response and the resonance mode in the superconducting state. Instead, the response is Y-shaped, gapped, and significantly enhanced below $T^*$, and hence a prominent signature of the pseudogap state.**

Most of the detailed knowledge about magnetic fluctuations in the cuprates comes from neutron scattering studies of $La_{2-x}Sr_xCuO_4$ (LSCO) and $YBa_2Cu_3O_{6+y}$ (YBCO)[1,2]. The momentum (**Q**) and energy ($\omega$) dependent dynamic magnetic susceptibility $\chi''(\mathbf{Q},\omega)$ of LSCO is characterized by an X-shaped hourglass spectrum that disperses with increasing energy from incommensurate wave-vectors at $\omega \approx 0$ toward the antiferromagnetic wave vector $\mathbf{q}_{AF}$, and then outward again at higher energies[1,3]. YBCO exhibits a similar dispersion in the superconducting (SC) state, yet its foremost characteristic is the magnetic resonance, a large and abrupt susceptibility increase at $\mathbf{q}_{AF}$ and a well-defined energy $\omega_r$ upon cooling below $T_c$ (ref. 2). Despite the apparent ubiquity of the hourglass dispersion, the differences between the spectra for LSCO and YBCO have motivated ostensibly contradictory microscopic interpretations: incommensurate spin-density-wave (SDW) fluctuations of local moments (unidirectional charge-spin stripe order is one example)[4], or a spin-exciton due to particle-hole excitations in the SC state[2,5]. The persistence of low-energy incommensurate excitations at temperatures well above $T_c$ for both LSCO and YBCO has stimulated speculation that the opening of the pseudogap (PG) is associated with SC[6] or SDW[1,7,8] fluctuations.

Although it has long been known that AF correlations may cause the *d*-wave superconductivity exhibited by the cuprates[9], such theoretical approaches were not thought to be able to account for the PG phenomenology. Following recent theoretical advancements[10], it has been argued that AF correlations may not only drive *d*-wave superconductivity, but potentially also PG electronic instabilities such as charge order (charge-density-wave (CDW), bond-density-wave), translational-symmetry-preserving ($\mathbf{q} = 0$) loop-current order, and pair-density-wave order[11-14]. These developments raise the prospect that much of the cuprate phase diagram may be understood as driven by AF correlations. It is therefore imperative to determine the detailed magnetic response in a structurally simple compound in which both CDW[15] and $\mathbf{q} = 0$ (refs. 16,17) order have been firmly established. Such measurements might also help illuminate the relevance of the seemingly universal hourglass response.

HgBa$_2$CuO$_{4+\delta}$ (Hg1201) features the highest optimal $T_c$ ($T_{c,max}$ = 97 K) of the single-CuO$_2$-layer cuprates (e.g., $T_{c,max}$ = 39 K for LSCO; YBCO has $T_{c,max}$ = 93 K and a double-CuO$_2$-layer structure), a simple tetragonal crystal structure, and minimal disorder effects[18]. Similar to the recent demonstration of the validity of Kohler's rule for the magnetoresistance in the PG phase[19], these model-system characteristics of Hg1201 can be expected to most clearly reveal the inherent magnetic fluctuation spectrum of the quintessential CuO$_2$ layers.

We present an inelastic neutron scattering study of the magnetic excitations of an underdoped Hg1201 sample with $T_c \approx$ 71 K and hole doping $p \approx$ 0.095 (labeled HgUD71; see Methods, Supplementary Notes 1-2 and Supplementary Figs. 1-5 for detailed experimental and analysis information). This is a particularly interesting doping level because it corresponds to the shoulder of the "SC dome", where $T_c$ appears to be slightly suppressed, and because a cascade of phenomena have been observed: quasi-static $\mathbf{q} = 0$ magnetic order[16,17] below $T^*$, short-range

CDW correlations[15] below $T_{CDW} \approx 200$ K, evidence for Fermi-liquid transport in the PG state[19-21], and Shubnikov-de Haas oscillations (below 4 K in magnetic fields above 60 T)[22]. The slight suppression of $T_c$ at this doping level might be a signature of a competing ground state, and it is interesting to determine if this has an effect on the dynamic magnetic susceptibility. As shown in Fig. 1a, this mirrors the phenomenology of underdoped YBCO[23-28]. We find that the dynamic magnetic response of HgUD71 is characterized by a gapped 'Y'-shaped dispersion that is commensurate with $\mathbf{q}_{AF}$ at energies below about 60 meV. Interestingly, the magnetic scattering exhibits a marked increase below the PG temperature $T^*$ yet is largely impervious to the onset of superconductivity. This establishes the commensurate excitations as a signature of the PG state.

**Results**

**Time-of-flight neutron spectroscopy.** Figure 2a,b shows $\chi''(\mathbf{Q},\omega)$ at 5 K and 85 K for HgUD71 extracted from raw neutron scattering data. At 5 K, deep in the SC phase, the susceptibility exhibits a prominent peak at $\omega_{peak} \approx 51$ meV and is gapped below $\Delta_{AF} \approx 27$ meV (see Supplementary Note 3 and Supplementary Fig. 7 for limits on the low-energy scattering). Up to approximately $\omega_{com} = 59$ meV, the response is commensurate with $\mathbf{q}_{AF}$, and then it disperses outward at higher energies, resulting in a gapped Y-shaped spectrum (Figs. 2a,c-g and 3e). The magnetic nature of the response is confirmed through spin-polarized neutron scattering and also from its **Q**-dependence, which follows the magnetic form factor (see Supplementary Note 4 and Supplementary Fig. 7).

We fit constant-**Q** data such as those in Figs. 2c-l to a Gaussian form, $\chi''(\mathbf{Q},\omega) = \chi_0'' \exp\{-4\ln 2\, R/(2\kappa)^2\}$, convolved with the momentum resolution of the instrument (Supplementary Fig. 2), where $R = |[(H-1/2)^2+(K-1/2)^2]^{1/2}-\delta|^2$, $2\kappa$ is the full-width-at-half-maximum (FWHM), and $\delta$ parameterizes the incommensurability away from $\mathbf{q}_{AF}$. The energy dependencies of $\chi_0''$, $2\kappa$ and $\delta$ are shown in Figs. 3a, 3b inset, and 3e, respectively.

**Commensurate low-energy magnetic excitations.** As shown in Fig. 2, the low-energy magnetic excitations in HgUD71 are commensurate with $\mathbf{q}_{AF}$. In Fig. 3e, $\delta = 0$ for $\omega < \omega_{com}$, since this results in the best fit to the data. We also fit the data with $\delta \neq 0$ (Fig. 4a) to facilitate comparison with published results for YBCO and LSCO (Figs. 4b,c), where at the neck of the hourglass a non-zero value of $\delta$ is typically employed in the data analysis even when the response is essentially commensurate. We find an upper bound of $\delta \approx 0.03$ r.l.u. for HgUD71 ($\omega < \omega_{com}$), which is consistent with the half-width-at-half-maximum (HWHM) of the instrumental momentum resolution (Fig. 4a and Supplementary Fig. 2). As seen from Fig. 4b,c, this upper bound is significantly smaller than the incommensurability observed in both LSCO and YBCO at similar doping levels.

**Evolution of magnetic excitations across $T^*$ and $T_c$.** Figure 3a shows the energy dependence of the susceptibility amplitude $\chi_0''$ at four temperatures. A comparison of the data at 5 K and 85 K reveals hardly any effect of superconductivity. This is also apparent from Fig. 3c, which shows the change $\Delta\chi_0''$ between these two temperatures. As seen from Fig. 3a, a further increase of temperature suppresses $\chi_0''$ at all measured energies (up to 53 meV for $T > 85$ K).

In order to better ascertain the temperature dependence of magnetic excitations, we focus on the response at $\omega_{peak} \approx 51$ meV and $\mathbf{q}_{AF}$ (Fig. 1b). Consistent with Fig. 3a, the intensity does not exhibit an abrupt change across $T_c$, which confirms the lack of a magnetic resonance. However, a marked increase in intensity occurs below the PG temperature $T^*$. For comparison, we measured the temperature dependence of the intensity of the odd-parity resonance mode of YBCO at a similar doping level (sample labeled YBCO6.6: $y = 0.6$, $T_c = 61$ K, $p = 0.11$, $\omega_r = 32$ meV; see black arrow in Fig. 1a). The result is shown in Fig. 1c. As for HgUD71, the intensity for YBCO6.6 increases below $T^*$, yet in contrast to HgUD71, a large magnetic resonance is observed below $T_c$.

Even though HgUD71 does not exhibit a magnetic resonance, upon cooling into the SC phase we observe subtle changes in the susceptibility at wave vectors away from $\mathbf{q}_{AF}$ and at energies below and above $\omega_{peak}$, centered at $\omega_1 = 44$ meV and $\omega_2 = 75$ meV. This is best seen from the momentum-integrated local susceptibility, $\chi''_{loc}(\omega) = \int \chi''(\mathbf{Q}, \omega) d^2q / \int d^2q$ (Fig. 3b,d). At $\omega_1$, the change across $T_c$ ($\Delta\chi''_{loc}$) is due to a slight increase of both the momentum width and intensity, whereas at $\omega_2$ it results from an increase in amplitude on the upward dispersive part of the spectrum (see also Fig. 2c,h,m). The $\mathbf{Q}$ dependence of these subtle changes across $T_c$ is discussed in more detail in Supplementary Note 5 and Supplementary Fig. 8.

**Discussion**

The absence of an hourglass dispersion in HgUD71 constitutes a clear departure from the purported universal magnetic response of the cuprates. Figure 4 compares the $\mathbf{Q}$-$\omega$ dispersion of the magnetic fluctuations centered at $\mathbf{q}_{AF}$ at similar doping levels for HgUD71, YBCO6.6 (refs. 7,8) and LSCO ($p \approx 0.085$, $T_c = 22$ K)[3]; see Supplementary Fig. 9 for an additional comparison between HgUD71 and LSCO ($p \approx 0.085$). LSCO exhibits a gapless hourglass dispersion both in the SC state and in the normal state. In the local-moment picture, the incommensurate low-energy response is argued to be a signature of SDW correlations[1]. For underdoped LSCO[1] and YBCO[28], the occurrence of incommensurate SDW order revealed by neutron scattering correlates with a planar resistivity characterized by a sizable extrapolated zero-temperature residual and by a low-temperature insulating-like upturn (when superconductivity is suppressed with large magnetic fields) below a non-universal critical doping $p_c$; $p_c \approx 0.16$ for LSCO[29] and $p_c \approx 0.085$ YBCO[30]. The doping-temperature range of the SDW correlations in YBCO is shown in Fig. 1a. At the doping level of our study, Hg1201 exhibits electrical transport without a significant zero-temperature residual, Kohler scaling of the normal-state magnetoresistance[19], as well as quantum oscillations[22], which demonstrates an underlying metallic ground state. Although $p_c$ for Hg1201 is not known, it is likely smaller than $p \approx 0.055$ ($T_c = 45$ K), for which the residual resistivity is still very small[22]. In addition to the *commensurate* low-energy response reported here for HgUD71, this indicates that Hg1201 is less prone to SDW order than YBCO and especially LSCO.

In the SC state, YBCO[7] (for $p > p_c$) exhibits a prominent resonance and a gapped magnetic response that is hourglass-shaped (Figs. 1c and 4b). The hourglass dispersion and resonance are best explained as signatures of the *d*-wave SC order parameter within the itinerant spin-exciton picture[2,5]. Although the resonance is well established over a wide doping range in double-layer YBCO[2], for single-layer compounds (Tl$_2$Ba$_2$CuO$_{6+\delta}$[2] and Hg1201[31]) it has been reported only close to optimal doping, where the PG phenomenon is relatively weak. According to the relationship $\omega_r/2\Delta_{SC} = 0.64 \pm 0.04$ found for unconventional superconductors[32], and with the estimate $2\Delta_{SC} = 78$ to 91 meV from electronic Raman spectroscopy[33] and photoemission[34], we expect $\omega_r \approx 54$ meV, which is close to $\omega_{peak} \approx 51$ meV for HgUD71. Interpreted within the spin-exciton picture, the suppression of the magnetic resonance at $q_{AF}$ might be due to an absence of coherent Bogoliubov quasiparticles at the 'hotspots' (where the underlying Fermi surface intersects the AF Brillouin zone boundary) as a result of the antinodal PG. This is consistent with electronic Raman spectroscopy for Hg1201[33,35], namely the fact that the SC pair-breaking peak in the $B_{1g}$ channel, which probes the antinodal states, significantly weakens upon underdoping (samples with $T_c$ below about 78 K), whereas the peak in the $B_{2g}$ channel, which probes the nodal states, persists. It is furthermore consistent with our observation of an increase of $\chi''(\mathbf{Q},\omega)$ below $T_c$ at momenta away from $\mathbf{q}_{AF}$ that connect parts of the Fermi surface closer to the coherent nodal directions that are unaffected by the PG.

The significant increase of the magnetic response below $T^*$ (Fig. 1b) and the concomitant absence of a prominent effect across $T_c$ indicates that the AF response for HgUD71 is dominated by the PG formation. The latter is a pivotal characteristic of the cuprates, and it is possibly associated with an underlying quantum critical point that controls much of the phase diagram[36,37]. A close connection between $\chi''(\mathbf{Q},\omega)$ and the PG has been suggested before[3,6,7]. In early work on YBCO, it was argued that the magnetic fluctuations in the PG state are a precursor of the resonance and therefore a signature of fluctuating superconductivity[6]. However, more recent work[7] on de-twinned YBCO6.6 found that $\chi''(\mathbf{Q},\omega)$ in the PG state is in fact distinct from that in the SC state. Furthermore, the broken four-fold structural symmetry of YBCO results in a large anisotropy in $\chi''(\mathbf{Q},\omega)$ for the two in-equivalent planar crystallographic directions. While the dispersion along [010] is reminiscent of the commensurate Y-shaped spectrum of HgUD71, the response along [100] is broader and incommensurate at low energies (Fig. 4c)[7]. This led to speculation that the PG is characterized either by stripe fluctuations, similar to LSCO, or by a nematic instability[7]. However, an alternative explanation for the anisotropy is interlayer coupling to the unidirectional CuO chain states of YBCO[38], a complication that is absent in HgUD71. Our result for this structurally simpler cuprate, showing a PG state characterized by a commensurate and isotropic low-energy magnetic response with no connection to a magnetic resonance, calls for a new theoretical interpretation. We speculate that the commensurate response for $\omega < \omega_{com}$ predominantly results from particle-hole scattering near the AF hotspots, and that it involves the non-dispersive region in the spectral density of states determined from scanning tunneling microscopy in the PG state[39].

A number of broken symmetries have been identified in the PG regime. In particular, the cuprates exhibit strong (~ 0.1 $\mu_B$) $\mathbf{q} = 0$ quasi-elastic magnetism[16,17,23,24] that is qualitatively consistent with intra-unit-cell loop-current order[37]. We demonstrate in Fig. 1b that the significant enhancement of fluctuations at $\mathbf{q}_{AF}$ coincides with the onset of $\mathbf{q} = 0$ magnetism for HgUD71, which establishes a connection between these two seemingly distinct magnetic phenomena and with the opening of the PG at $T^*$. On the other hand, (short-range) CDW correlations first appear at a temperature that is distinctly lower than $T^*$ (ref. 15) and have no discernible effect on the magnetic fluctuations (Fig. 1b,d, Supplementary Note 6, Supplementary Fig. 10). Regarding the changes across $T^*$, our result suggests that the development of AF correlations is a consequence rather than the cause of the PG. Nevertheless, these correlations might drive the subsequent CDW order, which in turn drives the Fermi-surface reconstruction implied by transport experiments in high magnetic fields[15,22,30]. It will be important to assess if this can indeed be the case given an instantaneous magnetic correlation length (estimated from integration over the measured energy range) of about 2 to 3 lattice constants in HgUD71.

Figures 1b-e show that, similar to HgUD71, for YBCO6.6 the intensity of the response at $\mathbf{q}_{AF}$ increases substantially along with the onset of $\mathbf{q} = 0$ order at $T^*$. Contrary to HgUD71, however, for YBCO6.6 this is followed by a large resonance below $T_c$ and by the concomitant appearance of the low-energy hourglass structure[7,8] and of a significant suppression of the CDW

response[26]. This highlights the differing relative strengths of the SC and PG order parameters at temperatures below $T_c$ for the two cuprates.

In summary, the AF response of the underdoped cuprates can be divided into three distinct types: (1) the gapless X-shaped spectrum associated with incommensurate SDW correlations of local moments in the La-based compounds[1] and in lightly-doped YBCO[28], where $q = 0$ magnetism is suppressed due to the competing SDW instability[24]; (2) the gapped X-shaped spectrum and magnetic resonance attributed to particle-hole excitations in the SC state[2,5]; and (3) the gapped Y-shaped spectrum associated with the PG formation (and with $q = 0$ magnetism[16,17] and metallic charge transport[19,20,22]) revealed most clearly in tetragonal Hg1201. The balance between PG, SDW, and SC order parameters determines the magnetic response for a particular compound, doping level and temperature. We note that similar to LSCO, single-layer $Bi_{2+x}Sr_{2-x}CuO_{6+y}$ (Bi2201) exhibits a propensity toward SDW order[40], whereas double-layer $Bi_2Sr_2CaCu_2O_{8+\delta}$ (Bi2212) near optimal doping features a dispersive resonance reminiscent of YBCO[41]. Just as for LSCO, $T_{c,max} = 38$ K[18] for Bi2201 is relatively low. Interestingly, the magnetic response of single-layer Hg1201 more closely resembles that of double-layer YBCO than those of single-layer LSCO and Bi2201. Yet the dominant PG behavior is most clearly apparent in Hg1201, which does not feature the complications of YBCO due to the orthorhombic double-layer structure (even vs. odd parity magnetic excitations; in-equivalent response along [100] and [010]). In order to build a connection with the distinct magnetic response of the low-$T_{c,max}$ single-layer compounds LSCO and Bi2201, it might be necessary to study Hg1201 with intentionally-introduced disorder. Furthermore, experiments on Hg1201 at lower doping levels will be necessary to ascertain if the SDW instability is in fact a universal property of the cuprates.

**Methods**

**Sample Preparation.** Single crystals of $HgBa_2CuO_{4+\delta}$ were grown by a two-step self-flux method[42]. As-grown crystals are typically underdoped, with $T_c \approx 81$ K. To reach the desired doping level, the crystals were annealed at 400°C in a partial vacuum of 100 mtorr for 80 days[43]. The superconducting transition temperature of the individual crystals was subsequently determined from measurements of the Meissner effect in a SQUID magnetometer: each crystal was cooled in zero magnetic field, and the susceptibility was monitored upon warming in a 5 Oe field applied along the crystallographic *c*-axis. Supplementary Fig. 1 shows the average susceptibility of all the 34 crystals that made up the HgUD71 sample. We find $T_c \approx 71$ K (defined as the mid-point of the transition) with a full transition width of $\Delta T_c = 5$ K for the assembled sample, and estimate $p \approx 0.095$ based on our thermoelectric power measurements of crystals from the same annealing batches. The doping level we estimate from these measurements is 0.005 higher than that estimated from prior published values for powder samples with the same $T_c$[44]. The rather narrow combined transition width indicates a high degree of homogeneity and quality of the sample. We note that smaller crystals from the same growth and annealing batches exhibit Shubnikov-de-Hass oscillations[22]. The fact that quantum oscillations can be observed at low temperatures is a consequence of the very small residual resistivity exhibited by these crystals[20]. The 34 crystals, with masses ranging from approximately 20 mg to 125 mg, were polished parallel to the *ab*-plane and co-aligned on two Aluminum plates with GE-varnish using a Laue backscattering X-ray machine. The resultant sample had a total mass of about 1.6 g and a planar mosaic of about 2 degrees. The plates were mounted on an Aluminum sample holder, as shown in Supplementary Fig. 1b. Gadolinium Oxide powder and Cadmium plates (both Ga and Cd are strong neutron absorbers) were used to mask the excess

Aluminum.

The YBCO6.6 sample (data in Fig. 1c) was previously measured in refs. 23 and 45. The sample was grown with a top-seed melt texturing method and heat-treated to an underdoped state with $T_c = 61 \pm 2.5$ K. We estimate the doping level to be $p = 0.11$ from the $T_c$ versus doping relation in ref. 46.

**Definition of wave vector.** We quote the scattering wave-vector $\mathbf{Q} = H\mathbf{a}^* + K\mathbf{b}^* + L\mathbf{c}^*$ as ($H$, $K$, $L$) in reciprocal lattice units (r.l.u.), where $a^* = b^* = 1.62$ Å$^{-1}$ and $c^* = 0.66$ Å$^{-1}$ are the room-temperature magnitudes. The reduced two-dimensional wave vector is $\mathbf{q} = h\mathbf{a}^* + k\mathbf{b}^*$, and $\mathbf{q}_{AF} = (1/2, 1/2)$.

**Time-of-flight measurements.** The time-of-flight (TOF) measurements were performed with the ARCS spectrometer at the Spallation Neutron Source, Oak Ridge National Laboratory. The HgUD71 sample was mounted such that the incoming beam was parallel to the $c$-axes of the sample. This means that for a particular in-plane wave vector ($H$, $K$), the out-of-plane component $L$ depends on the energy transfer. Two measurement configurations were used: incident energies $E_i = 70$ meV and 130 meV, with Fermi-chopper frequencies of 420 Hz and 600 Hz, respectively. The energy and momentum resolutions as a function of the energy transfer are presented in Supplementary Fig. 2. The out-of-plane wave-vector varies monotonically from $L \approx 2$ to 8 between $\omega = 10$ to 100 meV. As described in Supplementary Notes 1 and 2, the data are processed to isolate the AF fluctuations, and normalized by the magnetic form factor and Bose population factor to obtain $\chi''(\mathbf{Q},\omega)$. Inherent to our analysis is the assumption that the magnetic response arises from the quintessential CuO$_2$ planes and hence is quasi-two-dimensional, and that corrections for the $L$ dependence can be made by accounting for the Cu magnetic form factor.

**Triple-axis measurements with unpolarized neutrons.** Measurements on HgUD71 were performed with the HB3 spectrometer at the High-Flux Isotope Reactor at Oak Ridge National Laboratory (Fig. 1b). Measurements on YBCO6.6 (Fig. 1c) were performed with the 2T spectrometer at the Laboratoire Léon Brillouin (LLB, France) on the same twinned YBCO crystal ($T_c = 61 \pm 2.5$ K, $p = 0.11$) used to measure the **q** = 0 magnetic order in ref. 23. Pyrolytic graphite (PG) monochromators and analyzers were used to select incident and final neutron energies, and PG filters were used to suppress contamination due to higher harmonics. The samples were mounted in the (*HHL*) scattering plane. Measurements were performed with fixed final energies $E_f$ = 14.7 meV (HB3), and 35 meV (2T). On HB3, the horizontal collimation configuration was 48'-80'-sample-80'-120'. On 2T no collimation was used, since vertical and horizontal focusing was employed at the monochromator. The typical energy resolution in the $\omega$ = 50 − 60 meV energy transfer range was about 8 meV.

**Author Information**: Correspondence and requests for materials should be addressed to M.K.C. (mkchan@lanl.gov) and M.G. (greven@physics.umn.edu).

**Acknowledgements:** We acknowledge fruitful discussions with Yuan Li and Chandra Varma. We thank A. Kreyssig and A. I. Goldman, C. L. Broholm, and S. Koopayeh for assistance with crystal alignment work partially performed at Ames Laboratory and at the IQM at Johns Hopkins University. The work at the University of Minnesota was supported by the US Department of Energy, Office of Basic Energy Sciences, under Award No. DE-SC0006858. Research conducted at ORNL's High Flux Isotope Reactor and Spallation Neutron Source was sponsored by the Scientific User Facilities Division, Office of Basic Energy Sciences, US Department of Energy. M.K.C. is supported by funds from the US Department of Energy BES grant no. LANLF100. Work at the IQM was supported by U.S. Department of Energy, Office of Basic Energy Sciences, Division of Materials Sciences and Engineering under award DE-FG02-08ER46544. We also acknowledge financial support at LLB from the projects UNESCOS (contract ANR-14-CE05-0007) and NirvAna (contract ANR-14-OHRI-0010) of the ANR.

**Author Contributions**: M.K.C., L.M.T., Y.T. performed the neutron scattering experiments. M.K.C., C.J.D, Y.G., M.J.V., G.Y., and X.Z. performed crystal growth, characterization and co-alignment. A.D.C, J.T.P., Y.S., P.S., P.B., and D.L.A. were local contacts for the neutron scattering experiments. M.K.C., Y.S., P.B., and M.G. wrote the manuscript with input from all authors.

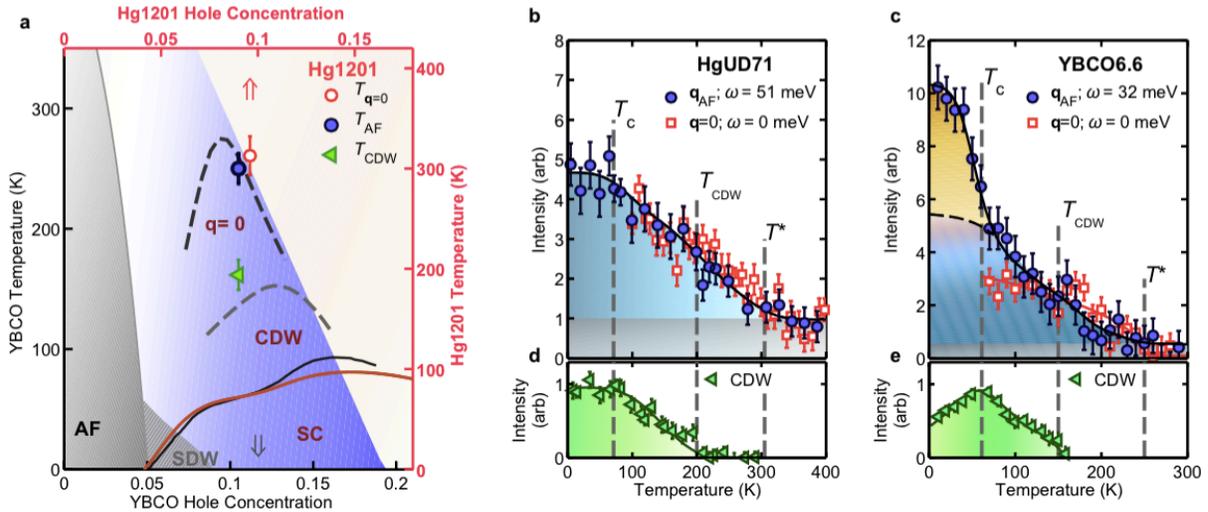

**Figure 1 | Phase diagram of Hg1201 and YBCO. (a)** Phase diagram: Hg1201 (red top-right axes) and YBCO (black bottom-left axes). The axes are adjusted such that the $T_c(p)$ domes of Hg1201 (red line) and YBCO (black line) approximately line up. The blue region represents the

PG regime. The red and grey arrows indicate the doping of HgUD71 ($T_c$ = 71 K) and YBCO6.6 ($T_c$ = 61 K) highlighted in this work. Dashed black and grey lines represent the temperatures below which $\mathbf{q}$ = 0 magnetic order[23,24] and CDW[25-27] correlations are observed in YBCO. The corresponding data for Hg1201 ($T_{\mathbf{q}=0}$, red circle; $T_{CDW}$, green triangle) for doping close to HgUD71 are shown[15-17]. The significant increase in the AF response appears below $T_{AF}$ (blue circle). Grey shaded and hashed areas represent AF and SDW[28] order in YBCO. (**b**) Temperature dependence of inelastic magnetic scattering at $\mathbf{q}_{AF}$ and $\omega_{peak}$ = 51 meV and of quasi-elastic (FWHM energy resolution ~1 meV) magnetic scattering at $\mathbf{q}$ = 0 (from ref. 17) measured on separate Hg1201 samples with similar doping levels ($T_c$ = 71 K and 75 K, respectively). We determine $T_{\mathbf{q}=0}$ = 320 ± 20 K (ref. 17) and $T_{AF}$ = 300 ± 15 K. The $\mathbf{q}$ = 0 signal is shifted upward (by 1 unit) for comparison. Both $T_{\mathbf{q}=0}$ and $T_{AF}$ are consistent with $T^*$ = 305 ± 10 K determined from the planar resistivity (deviation from high-temperature linear dependence)[19]. (**c**) Temperature dependence of the $\mathbf{q}$ = 0 signal[23] and the odd-parity resonance energy ($\omega_r$ = 32 meV) for a twinned YBCO6.6 sample (see Methods for sample details). For **b**&**c**, the blue region represents the increase in intensity in the PG state and the grey region is the baseline intensity for $T > T^*$. The shaded orange region in **c** represents the excess scattering below $T_c$ due to the resonance mode. (**d**&**e**) Temperature dependence of short-range CDW order in Hg1201 (ref. 15) and YBCO (ref. 26) at approximately the same respective doping levels as the data in **b**&**c**. Vertical error bars in **a**,**b**&**c** are statistical errors (1 s.d.).

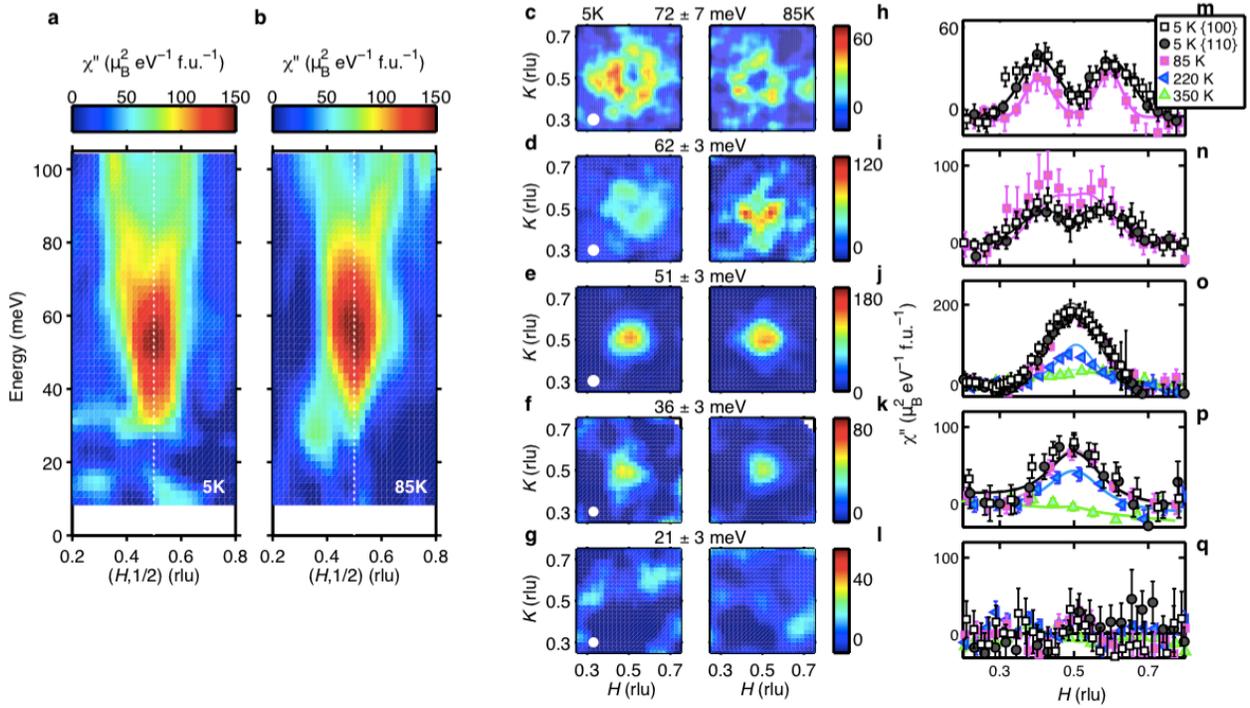

**Figure 2 | Magnetic excitation spectrum of HgUD71 features gapped, commensurate and dispersive components.** (**a**,**b**) Energy dependence of $\chi''(\mathbf{Q},\omega)$ for HgUD71 at 5 K and 85 K (14 K above $T_c$), respectively, along the two-dimensional momentum-transfer trajectory [$H$,0.5], with intensity averaged over the range $K = 0.5 \pm 0.12$. The gap $\Delta_{AF}$ is defined as the energy below which no scattering is observed at $\mathbf{q}_{AF}$; $\Delta_{AF} \approx 27$ meV at both 5 K and 85 K. (**c-l**) Constant-energy slices of magnetic scattering at $T = 5$ K in **c-g** and $T = 85$ K in **h-l**. (**m-q**) The corresponding constant-energy cuts along high-symmetry trajectories. Cuts along [100] and [010] are shown for $T = 5$ K (open and closed black squares, respectively). Data at higher temperatures (85 K, 220 K and 350 K) are averages of cuts along {100} and {010} trajectories. Error bars represent statistical error (1 s.d.). The white circles in **c-g** represent the momentum resolution at the corresponding energy transfers. Data collected on ARCS (see Methods).

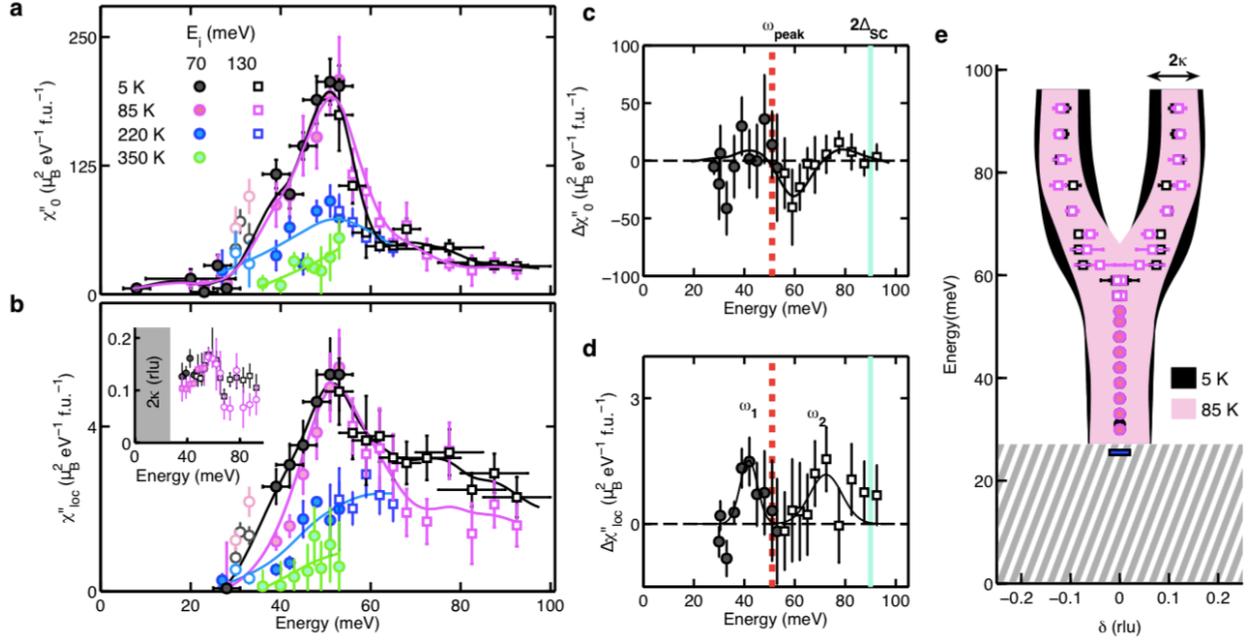

**Figure 3 | Magnetic susceptibility amplitude and local susceptibility for HgUD71. (a)** Energy dependence of the measured peak magnetic susceptibility $\chi''_0$ at $T$ = 300 K, 220 K, 85 K and 5 K. Closed circles: $E_i$ = 70 meV. Open squares: $E_i$ = 130 meV. Solid lines: guides to the eye. Horizontal bars for the 5 K data represent energy bins. The same binning is used at higher temperatures. Between 30 meV and 33 meV the data are systematically contaminated by Aluminum and phonon scattering, and are represented as lighter open symbols (see Supplementary Fig. 3). **(b)** Same legend as **a**. Energy dependence of the momentum-integrated (local) susceptibility $\chi''_{loc}$. In determining $\chi''_{loc}$, we assume that AF fluctuations are quasi-two-dimensional, i.e., that $\chi''$ does not depend on $L$. **Inset:** $2\kappa$ (FWHM) as a function of energy at 5 K (black) and 85 K (magenta). **(c,d)** Change of $\chi''_0$ and $\chi''_{loc}$, respectively, between 5 K and 85 K (i.e., across $T_c$). Filled and open symbols: $E_i$ = 70 meV and 130 meV, respectively. The red vertical line marks $\omega_{peak}$. The turquoise line represents $2\Delta_{SC}$, where $\Delta_{SC}$ = 45 ± 1 meV is the maximum SC $d$-wave gap determined from Raman scattering[33]. Black line in **c**: guide to the eye. Black line in **d**: fit to two Gaussian peaks, located at $\omega_1$ = 44 ± 2 meV and $\omega_2$ = 75 ± 2 meV. **(e)**

Energy dependence of incommensurability δ at 5 K (black) and 85 K (red). Horizontal error bars are fit uncertainties for δ. For $\omega < 59$ meV, the data are best described with $\delta = 0$. We estimate an upper bound of $\delta \approx 0.03$, which is the approximate value of the instrumental momentum resolution in the $\omega = 27 - 59$ meV range. Shaded black and magenta regions represent $2\kappa$ at 5 K and 85 K, respectively. Hatched area indicates the gap $\Delta_{AF}$. Horizontal blue bar at $\omega = 27$ meV represents the instrumental momentum resolution at that energy for $E_i = 70$ meV. All vertical error bars in figure are least-square fit errors (1 s.d.).

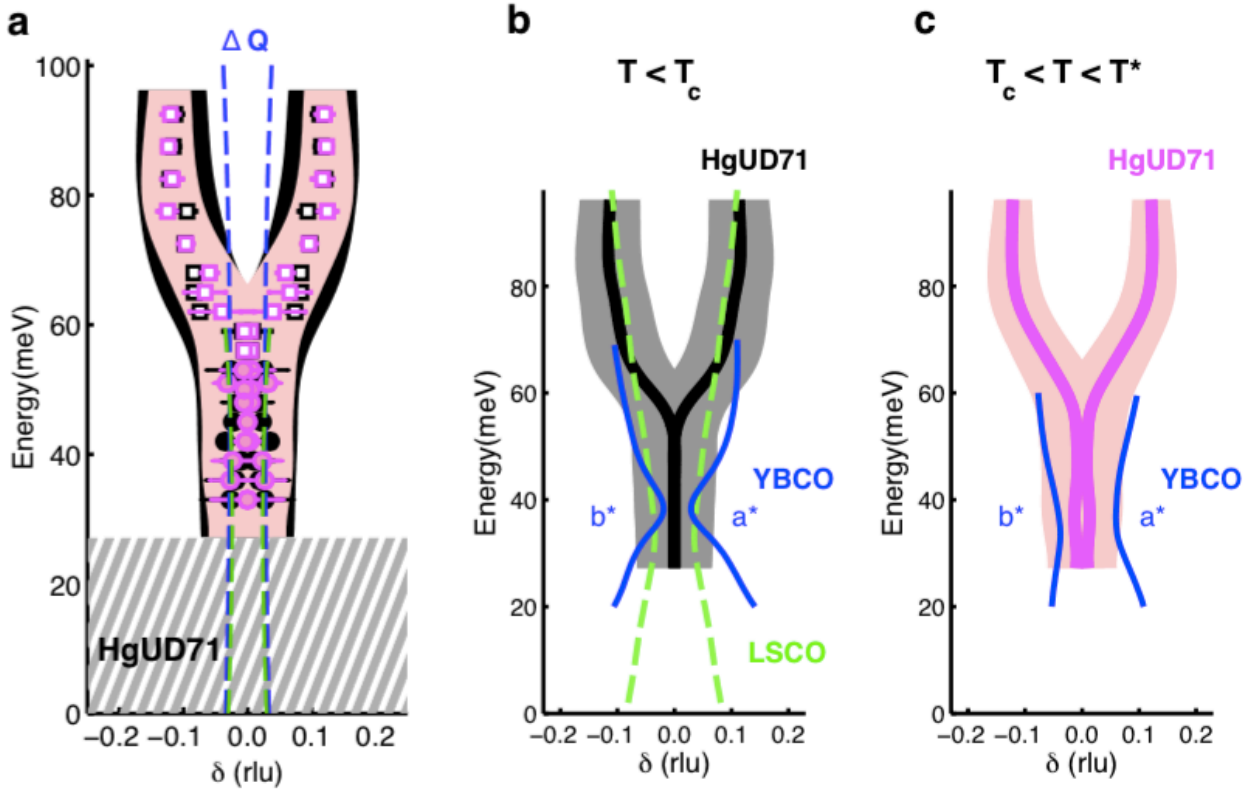

**Figure 4 | Comparison of Y-shaped response of HgUD71 with magnetic excitation spectrum of YBCO and LSCO at similar doping levels.** (**a**) Energy dependence of incommensurability $\delta$ at 5 K (black) and 85 K (magenta), with $\delta$ assumed to be non-zero at all energies. We arrive at an upper bound of $\delta \sim 0.03$ r.l.u., which corresponds to the HWHM of the instrumental **Q** resolution

(FWHM resolution for $E_i$ = 70 meV and 130 meV indicated by the green and blue dotted lines, respectively). Shaded black and red regions represent the measured FWHM ($2\kappa$) at 5 K and 85 K, respectively. Hatched grey area indicates the gap in the excitation spectrum. (**b**) Comparison of the dispersion of HgUD71 with YBCO6.6 (ref. 7,8) and LSCO ($p = x = 0.085$)[3] in deep the SC state. (**c**) Comparison of the dispersion of HgUD71 with YBCO6.6 above $T_c$ ($T$ = 85 K and 70 K, respectively)[7]. The response of orthorhombic YBCO6.6 (blue lines) is anisotropic, and therefore $\delta$ along both **a*** (right) and **b*** (left) is shown. As in **a**, the shaded regions in **b&c** indicate the momentum widths (FWHM) of the response of HgUD71.

# Supplementary Information:

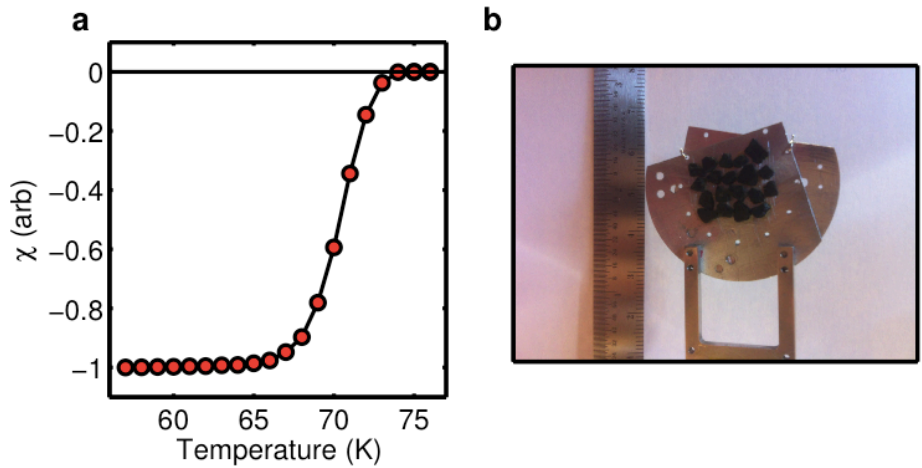

**Supplementary Figure 1 | Sample HgUD71.** (**a**) Average magnetic susceptibility of the constituent 34 crystals of sample HgUD71 with mid-point transition temperature $T_c$ = 71 K. (**b**) Picture of the two Aluminum plates with mounted crystals.

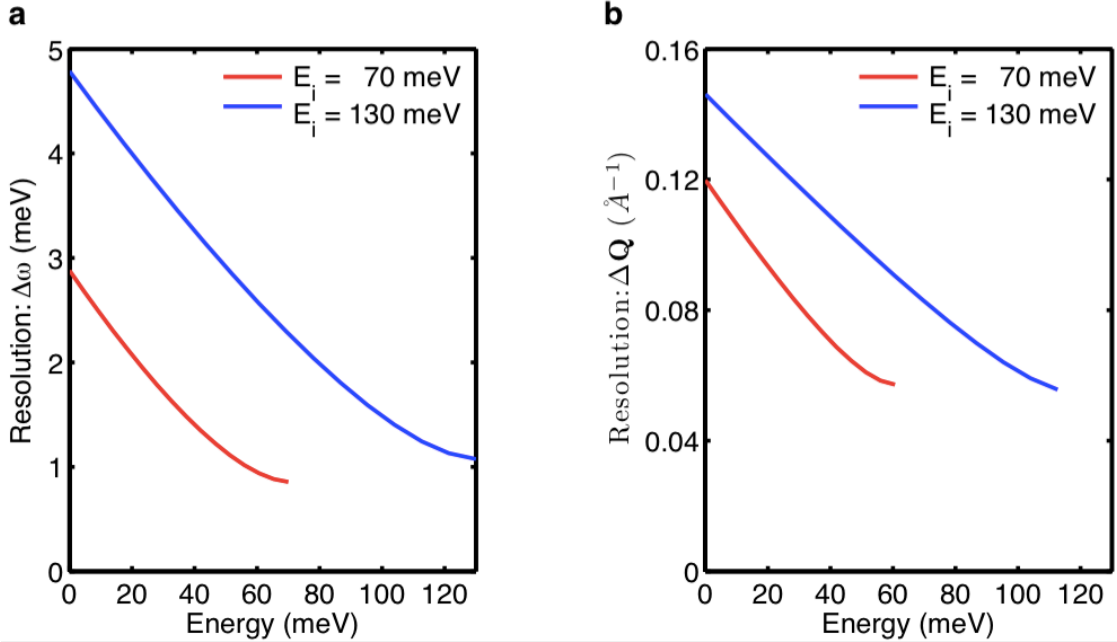

**Supplementary Figure 2 | ARCS instrumental resolution.** (**a**) Full-width-at-half-maximum (FWHM) energy resolution $\Delta\omega$ as a function of the energy transfer, determined by considering the contributions to the timing uncertainty from the source and chopper openings in conjunction with the neutron flight-path lengths[1]. The energy resolution improves as the energy transfer approaches the incident energy, $E_i$. (**b**) FWHM momentum resolution determined at the two-dimensional antiferromagnetic wave vector $\mathbf{q}_{AF}$ = (1/2, 1/2). Note that 1 r.l.u. = $2\pi/a \approx 1.63$ Å$^{-1}$, so that 0.08 Å$^{-1}$ ≈ 0.05 r.l.u. A sample mosaic of 2° was considered in the determination of the momentum resolution. The calculated energy and momentum resolutions were confirmed by comparing to the measured widths of elastic Bragg peaks and incoherent scattering.

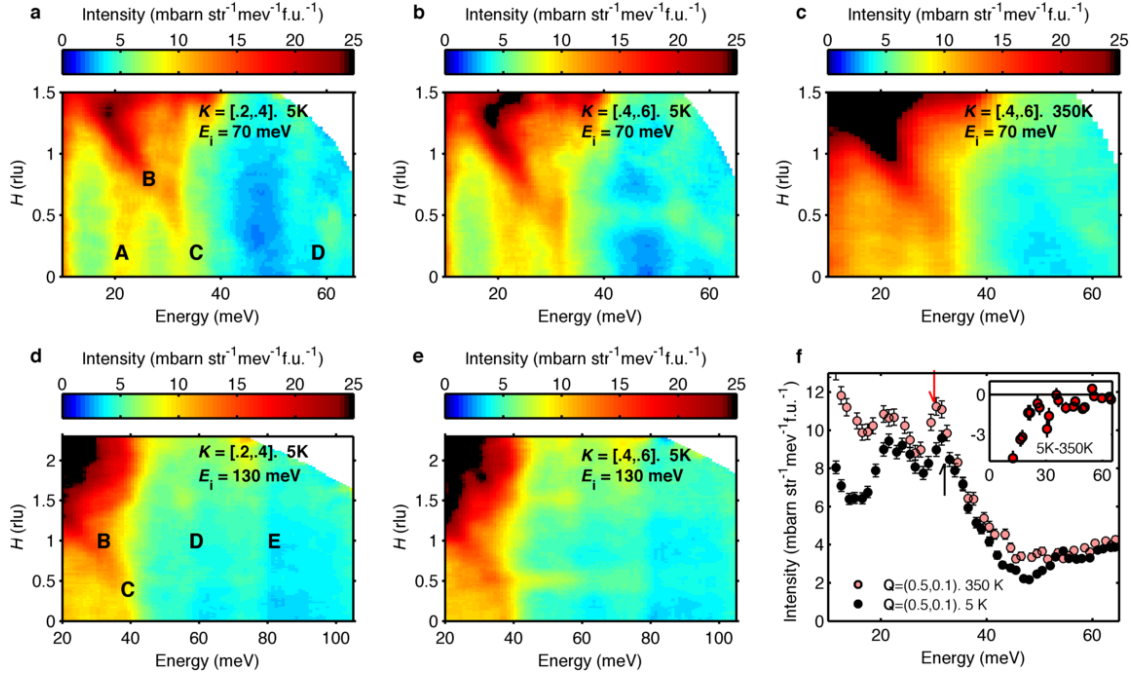

**Supplementary Figure 3 | Additional contributions to the measured intensity.** (**a-b**) Energy dependence of total scattering intensity at $T = 5$ K, with $E_i = 70$ meV, along the momentum transfer trajectories [$H$, 0.3±0.1] and [$H$, 0.5±0.1], respectively. The letters A-D in **a** mark various contributions to the scattering besides the AF fluctuations that we are concerned with: A and C are optical phonons with nearly zero dispersion; B is a dispersing phonon from a powder contribution, probably Aluminum, and presents as a ring of scattering in constant-energy slices, as demonstrated in Supplementary Fig. 5; D is one of the weakly-dispersing modes reported previously for the pseudogap phase of Hg1201 (Supplementary refs. 5 and 6). In **b**, AF fluctuations centered at $\mathbf{q}_{AF} = (0.5, 0.5)$ are clearly seen in the raw data superimposed on the "background" contributions A-D. As demonstrated in Fig. 2 of the main text, the AF fluctuations can be successfully isolated by subtracting the various additional contributions. However the scattering within the energy range 30-36 meV around $\mathbf{Q}_{AF}$ is typically highly contaminated by the Aluminum and phonon lines, which results in a slight overestimation of the magnetic

contribution. (**c**) Same as **b**, but at 350 K, above the pseudogap temperature. The contributions A-C are more intense, whereas D is no longer observed. This confirms the identification of A-C as phonons and D as a possible magnetic mode[5-6]. (**d-e**) Same as **a&b**, but with $E_i$ = 130 meV. The features B-D can be identified. An additional dispersionless feature is observed at ~ 75 meV, marked by E. Although high-temperature data were not taken with $E_i$ = 130 meV, unpublished measurements at 420 K on a more underdoped Hg1201 sample ($T_c \approx$ 55 K, $T^*$ = 400 K) show that, unlike the mode at 55 meV (feature D) this dispersionless feature is still observed at high temperatures. The origin of this feature is undetermined. (**f**) Energy dependence of scattering at **q** = (0.5, 0.1) for 5 K and 350 K, with $E_i$ = 70 meV. The 30 meV phonon (marked by the red and black dashed lines) softens by about 1-2 meV with increasing temperature.

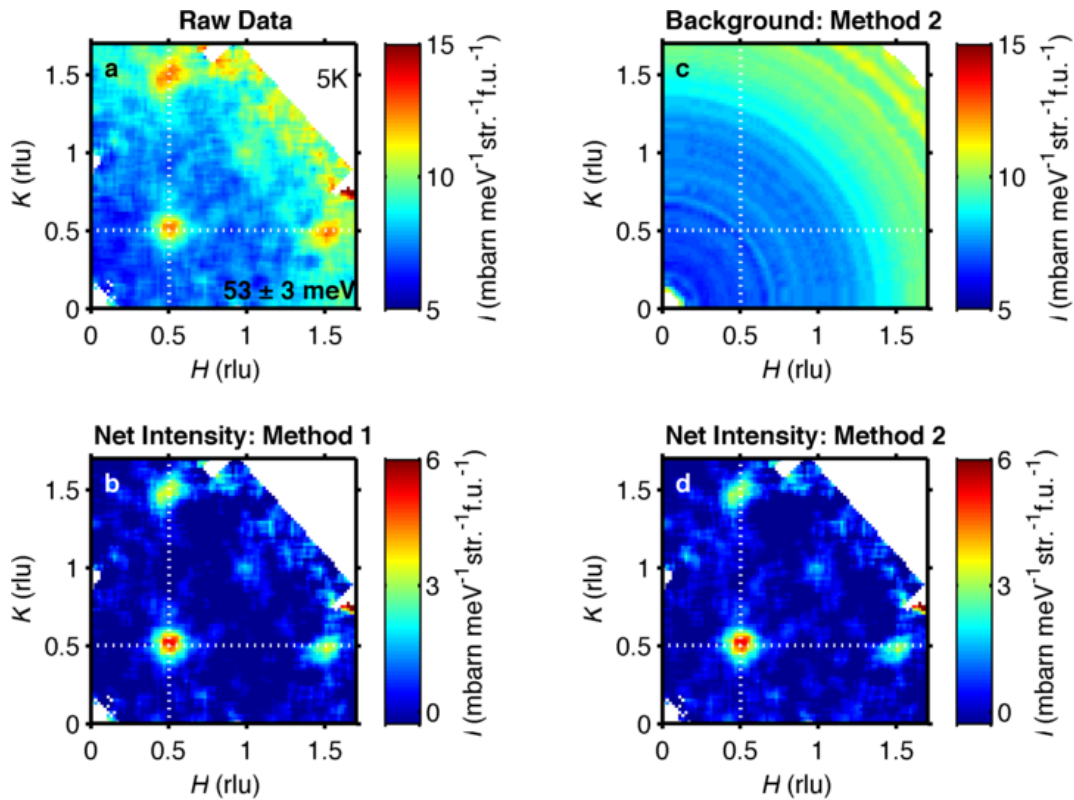

**Supplementary Figure 4 | Data processing – Part I.** (**a**) Representative constant-energy slice ($\omega = 53 \pm 3$ meV) of the raw data. In order to isolate the AF fluctuations, two possible methods are employed. (**b**) Result of implementing Method 1. (**c**) Calculated "background" level determined from Method 2. (**d**) Resultant net intensity with Method 2. For this particular energy slice, both methods produce essentially equivalent results. As demonstrated in Supplementary Fig. 5, Method 2 must be used when spurious contributions from powder lines are present.

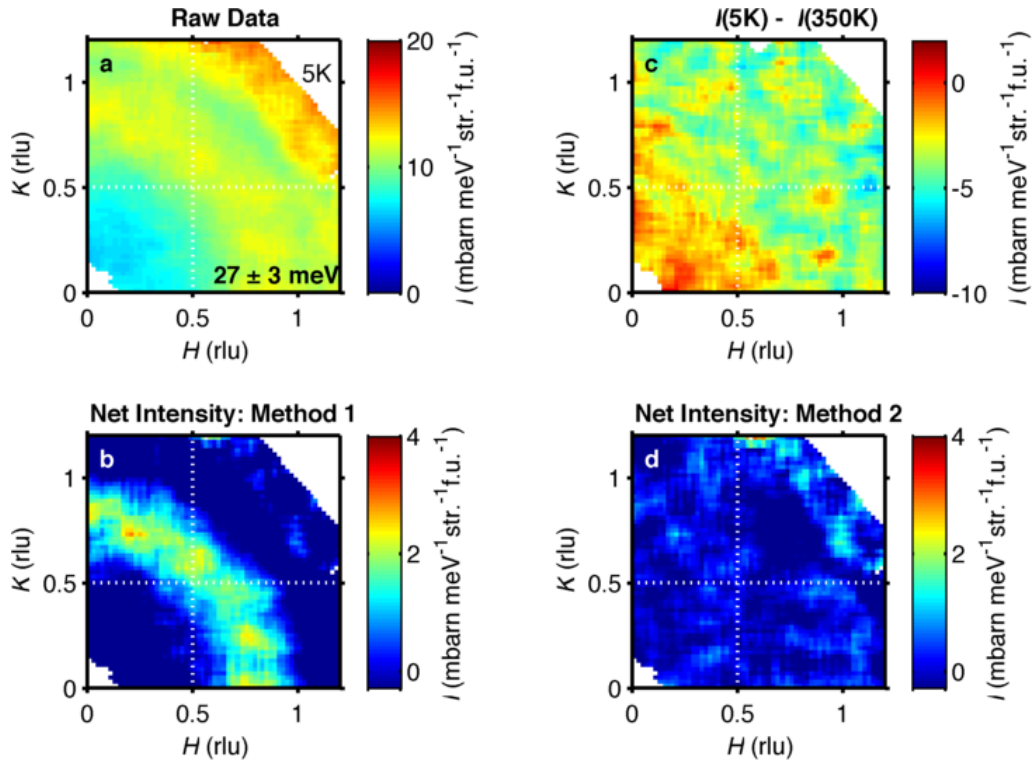

**Supplementary Figure 5 | Data processing – Part II.** (**a**) Example of a constant-energy slice ($\omega = 27 \pm 3$ meV) in which a powder ring obscures the underlying behavior at $\mathbf{q}_{AF}$. (**b**) Net intensity after subtracting the "background" contributions determined using Method 1. It is clear that this method does not adequately remove the spurious powder ring. (**c**) The non-AF contributions visible in the raw data can be largely removed by taking the temperature difference between data at 5K and 350 K. (**d**) Method 2 successfully removes the powder ring without taking a temperature difference. Consistent with the discussion in the main text, no AF fluctuations are observed for $\omega \approx 27$ meV.

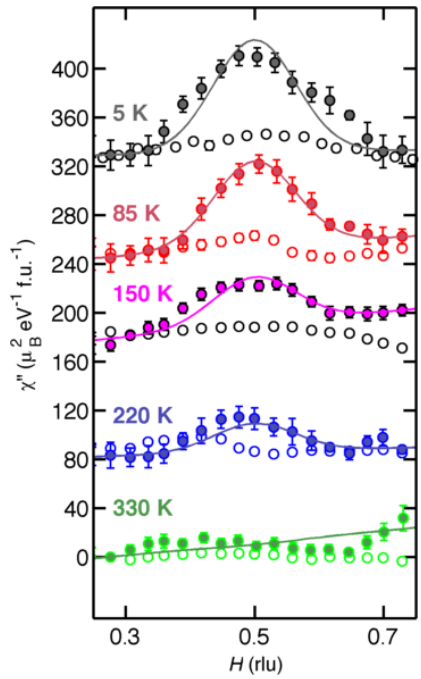

**Supplementary Figure 6 | Temperature dependence of the low-energy response.** Temperature dependence of the magnetic susceptibility at $\omega = 36 \pm 3$ meV (solid symbols) and $15 \pm 5$ meV (open symbols). Cuts along along {100} and {010} are averaged. Solid lines are Gaussian fits to the $\omega = 36$ meV data.

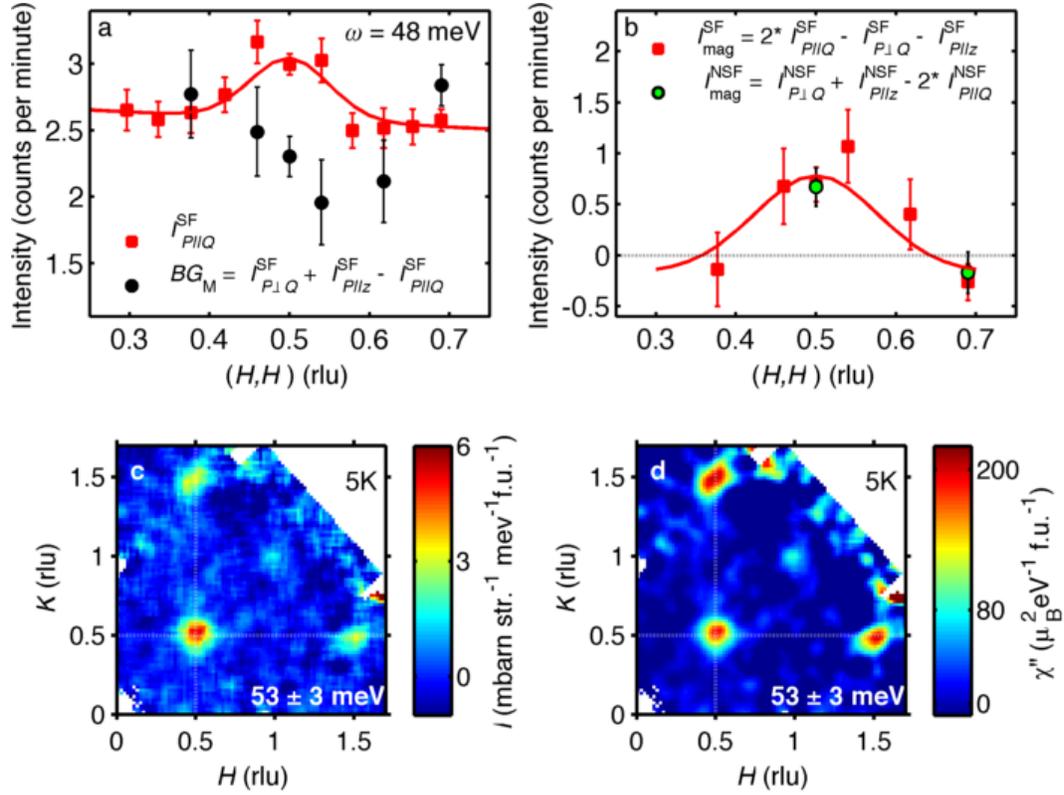

**Supplementary Figure 7 | Verification of magnetic scattering.** (**a**) Rocking-scan of spin-flip (SF) scattering about **Q** = (0.5, 0.5, 3.5) at $\omega$ = 48 meV and $T$ = 2 K (red squares) for initial neutron spin polarization **P** parallel to **Q** ($I^{SF}_{P\|Q}$). SF scattering measures magnetic fluctuations perpendicular to both **Q** and **P**, as well as incoherent nuclear spin scattering and background contributions. The background level for magnetic scattering defined as $BG_M \equiv \left(\frac{2}{3}N_{\text{inc,spin}} + BG\right)$ is determined in longitudinal polarization analysis (black circles) through additional measurements in the two other principal geometries ($I^{SF}_{P\|z}$ and $I^{SF}_{P\perp Q}$). Excess scattering above the background level, which we attribute to magnetic fluctuations, is clearly observed at $\mathbf{q}_{AF}$. (**b**) The intensity measured in the three SF geometries can be used to extract the pure magnetic scattering (red squares). This is confirmed through a corresponding NSF measurement (green circles). Error bars in **a** and **b** represent statistical uncertainty (1 s.d.). (**c**) Constant-energy ($\omega$ = 53 ± 3

meV) slice from TOF measurement showing magnetic scattering intensity in several Brillouin zones. The distinct peak at **Q** = (0.5, 0.5, 4.55) is repeated at **Q** = (1.5, 0.5, 6.34) and (0.5, 1.5, 6.34). (**d**) The data from **c** are normalized by the anisotropic magnetic form factor for Cu $3d_{x^2-y^2}$ (ref. 2) to obtain $\chi''$.

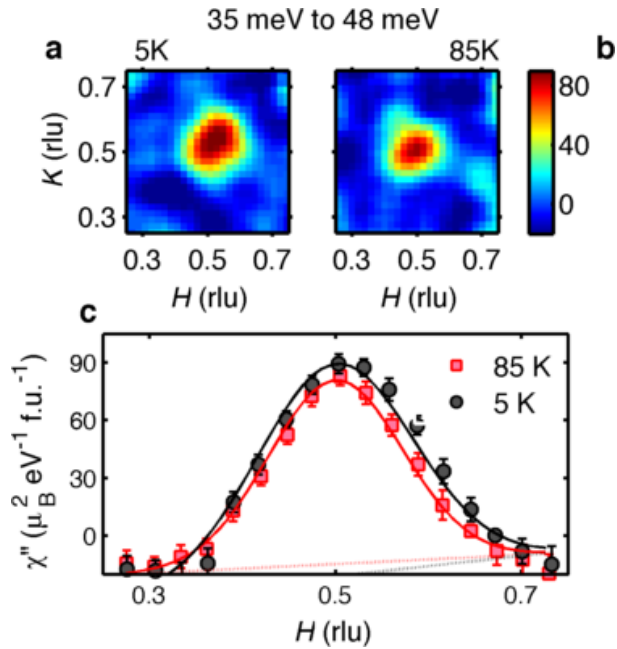

**Supplementary Figure 8 | Effect of superconductivity on the width of the low-energy response.** (**a**),(**b**) $\chi''(\mathbf{Q})$ constant energy slice integrated between 35 meV and 48 meV at 5 K and 85 K respectively. (**c**) Corresponding cut of the data averaged over the {100} and {110} trajectories of the data in (a) and (b). The peak at 5 K is slightly broader than that at 85 K, which accounts for the enhancement at $\omega_1$ away from $\mathbf{q}_{AF}$ in $\Delta\chi'' = \chi''(5\ K) - \chi''(85\ K)$ discussed in Fig. 3d.

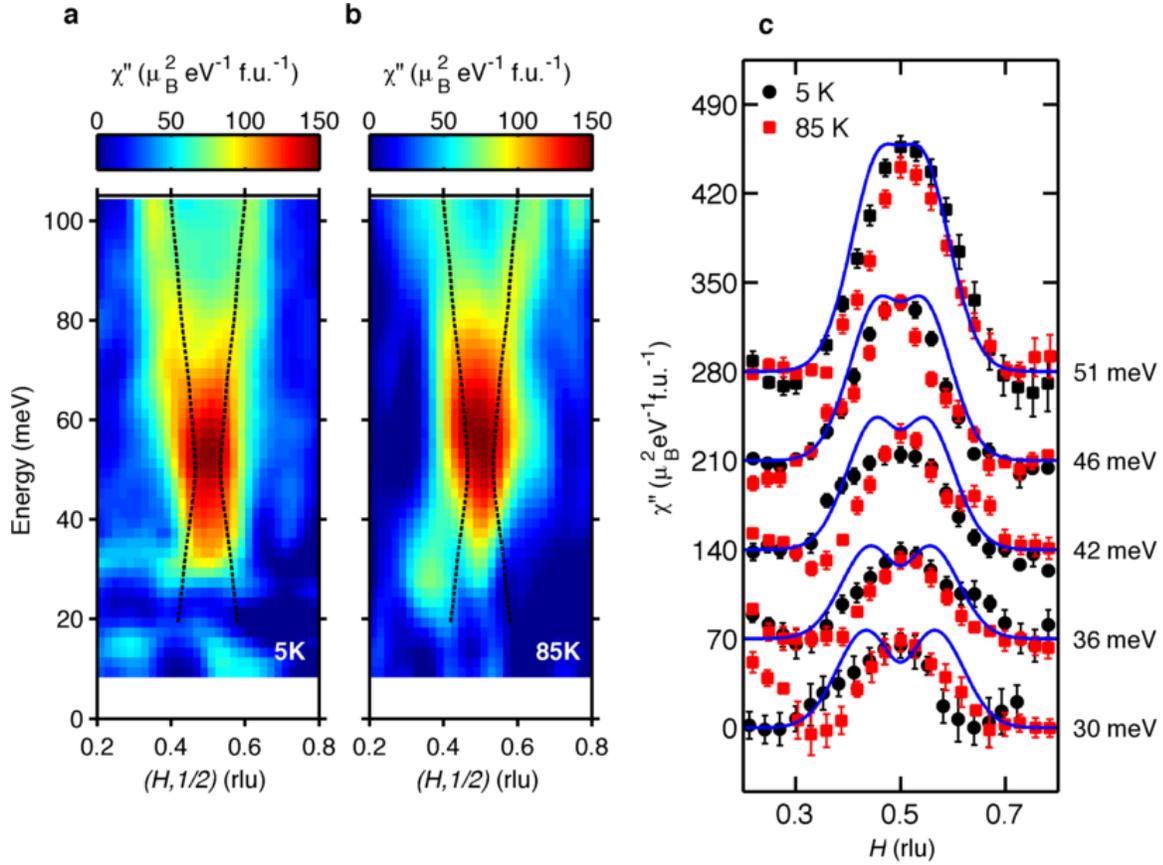

**Supplementary Figure 9 | Comparison to LSCO (a-b)** Energy dependence of $\chi''(\mathbf{Q},\omega)$ for HgUD71 at 5 K and 85 K along the two-dimensional momentum-transfer trajectory $[H,0.5]$ with $E_i = 130$ meV, reproduced from Fig. 2 of the main text. Dotted black lines show the dispersion of LSCO ($p = 0.085$)[8] shifted to higher energies by 17 meV such that the neck of the hourglass coincides with the peak in magnetic scattering at 51 meV for HgUD71, as expected based on the phenomenology of the cuprates[9,10]. Whereas the high-energy ($\omega > 50$ meV) dispersion for HgUD71 is similar to that for LSCO, this is not the case for the low-energy response, which exhibits an outward dispersion for LSCO and is commensurate for HgUD71. **(c)** Constant-energy cuts with {100} and {010} trajectories averaged to improve statistics. Due to the better counting statistics, data obtained with $E_i = 70$ meV rather than $E_i = 130$ meV are shown. The data are offset vertically for clarity. The blue lines show the hypothetical response for LSCO for the

hourglass dispersion away from the neck indicated in (a) and (b) with FWHM = 0.12 rlu (consistent with our fitting) and convolved with our experimental resolution for $E_i$ = 70meV . The FWHM was chosen to approximate that determined for HgUD71 in the relevant energy range (inset to Fig. 3b in the main text). It is clear the low-energy response of HgUD71 is not consistent with the dispersion observed for LSCO.

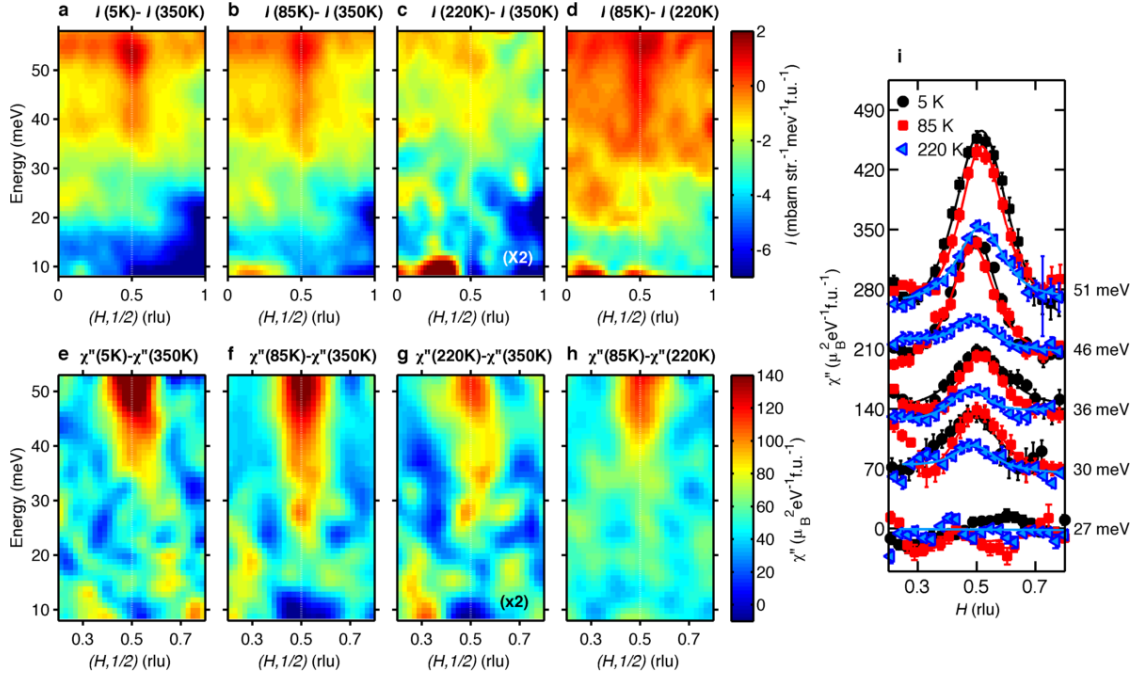

**Supplementary Figure 10 | Temperature dependence of the AF response.** (**a-d**) Raw scattering intensity at $T$ = 5 K, 85 K and 220 K, respectively, after subtracting the data taken at 350 K (shown in Supplementary Fig. 3c). Note that 350 K is above $T^*$. Data taken with $E_i$ = 70 meV. The intensity in **c** is multiplied by a factor of two to highlight weak features. Both the AF fluctuations and dispersionless modes at about 40 meV and 55 meV are visible. The two Ising-like modes were recently reported for optimally-doped ($T_c \approx$ 95 K) and under-doped ($T_c \approx$ 65 K) Hg1201[5-6]. (**e-g**) $\chi''(\mathbf{Q},\omega)$ determined by processing the data in **a-c** with Method 2, and converting to the susceptibility. The dispersionless features are absent because they are explicitly removed by this procedure, which isolates the AF fluctuations. (**d,h**) Same as above, but with focus on the change of the response between 220 K (just above the CDW temperature) and 85 K (just above $T_c$). (**i**) Line cuts of the data across $\mathbf{q}_{AF}$ after background subtraction as in panels **e-g**. Cuts along {100} and {010} trajectories were averaged to improve statistics. Data at each energy are offset vertically for clarity.

**Supplementary Note 1**

**Analysis of time-of-flight neutron scattering data.** Supplementary Figs. 1 and 2 provide sample and instrument resolution information, respectively. The energy and momentum dependence of the raw measured scattering intensities are shown in Supplementary Fig. 3. Besides the AF fluctuations, manifest as rods of scattering centered at $\mathbf{q}_{AF}$, a number of additional features are observed. These are identified in Supplementary Figs. 3a and 3d. In order to properly analyze the AF fluctuations they have to be isolated from these other contributions. Supplementary Figs. 4 and 5 demonstrate the use of two methods to subtract an energy ($\omega$) and wave-vector ($\mathbf{Q}$) dependent "background" $BG(\omega,\mathbf{Q})$. Method 1 fits a constant-energy slice to a second-order polynomial in the magnitude of the wave vector: $BG(\omega,\mathbf{Q}) = A(\omega) + B(\omega)|\mathbf{Q}| + C(\omega)\mathbf{Q}^2$. In Method 2, $BG(\omega,\mathbf{Q})$ is taken to be the average of the intensity at a given wave-vector magnitude $|\mathbf{Q}|$ at fixed energy transfer. In both methods, the data around $\mathbf{q}_{AF}$ are excluded from the determination of the additional contributions. Although Method 1 is more straightforward to implement, Method 2 is required to remove scattering from powder lines. This is evident from the comparisons in Supplementary Figs. 4 and 5. The processed data presented in the main text are based on Method 2 and, where feasible, the results are confirmed with Method 1.

**Supplementary Note 2**

**Obtaining absolute units of the dynamic magnetic susceptibility**. TOF data are converted to absolute units by comparison with a Vanadium standard. Assuming isotropic magnetic fluctuations, the AF scattering cross-section is related to the imaginary part of the dynamic magnetic susceptibility, $\chi''(\mathbf{Q},\omega)$, via[2]

$$\frac{d^2\sigma}{d\Omega dE} = \frac{2(\gamma r_e)^2}{\pi g^2 \mu_B^2} \frac{k_f}{k_i} |F(\mathbf{Q})|^2 \frac{\chi''(\mathbf{Q},\omega)}{1 - \exp(-\omega/k_B T)}$$

where $(gr_e)^2 = 0.2905$ barn sr$^{-1}$, $k_f$ and $k_i$ are the final and incident neutron wave-vectors, and $|F(\mathbf{Q})|^2$ is the magnetic form factor. Supplementary Fig. 6c shows a constant energy image with peaks of intensity at $\mathbf{q}_{AF}$ at $\omega = 53$ meV. The scattering intensity is larger at $\mathbf{Q} = (0.5, 0.5, 4.55)$ than in higher (two-dimensional) Brillouin zones ($\mathbf{Q} = (0.5, 1.5, 6.34)$ and $(1.5, 0.5, 6.34)$), consistent with magnetic scattering, for which the form-factor is smaller at larger $\mathbf{Q}$. After accounting for the anisotropic Cu$^{2+}$ $d_{x^2-y^2}$ form factor[2], $\chi''$ becomes equivalent in all zones (Supplementary Fig. 6d).

**Supplementary Note 3**

**Limits on the low-energy response.** Supplementary Fig. 6 shows the temperature dependence of response at $\omega = 15 \pm 5$ meV and $\omega = 36 \pm 3$ meV. There is no discernible magnetic scattering above the concave background at $\omega = 15$ meV, and we estimate conservative upper bounds of 15% and 8%, respectively, compared to the signal at $\omega = 36$ meV and 51 meV. We note that $\chi''_0(\omega)$ for $\omega > \Delta_{AF}$ extrapolates to zero at $\omega \sim 25$ meV (Fig. 3a; 5 K data), consistent with $\Delta_{AF} = 27$ meV, which is defined as the energy below which we can no longer observe a peak at $\mathbf{Q}_{AF}$.

**Supplementary Note 4**

**Polarized neutron scattering.** Spin-polarized measurements (Supplementary Fig. 7a,b) were carried out on the triple-axis spectrometer IN20 at the Institute Laue Langevin. Heusler alloy crystals were used as monochromator and analyzer to select the initial and final neutron energies and spin polarizations. The polarization of the neutron beam in the vicinity of the sample was maintained by CryoPAD, which provides high stability and reproducibility of the neutron spin polarization[4].

For longitudinal polarization analysis, it is convenient to define the coordinate system based on the relative orientations of the neutron spin polarization (**P**) at the sample, the scattering wave-vector **Q**, and the scattering plane that contains **Q**: the three orthogonal axes are defined by **P**∥**Q** and **P**⊥**Q** in the scattering plane, and **P**∥**z**, where **z** is the direction perpendicular to the scattering plane. In the absence of chiral magnetic correlations, the measured spin-flip (SF) and non-spin-flip (NSF) scattering intensities in the three principal spin-polarization geometries are given by the following relations[1]:

$$I^{SF}_{P\|Q} = \frac{2}{3}N_{\text{inc,spin}} + M_{P\perp Q} + M_{P\|z} + BG \tag{1}$$

$$I^{NSF}_{P\|Q} = N_{\text{coh}} + N_{\text{inc,isotope}} + \frac{1}{3}N_{\text{inc,spin}} + BG \tag{2}$$

$$I^{SF}_{P\perp Q} = \frac{2}{3}N_{\text{inc,spin}} + M_{P\|z} + BG \tag{3}$$

$$I^{NSF}_{P\perp Q} = N_{\text{coh}} + N_{\text{inc,isotope}} + \frac{1}{3}N_{\text{inc,spin}} + M_{P\perp Q} + BG \tag{4}$$

$$I^{SF}_{P\|z} = \frac{2}{3}N_{\text{inc,spin}} + M_{P\perp Q} + BG \tag{5}$$

$$I^{NSF}_{P\|z} = N_{\text{coh}} + N_{\text{inc,isotope}} + \frac{1}{3}N_{\text{inc,spin}} + M_{P\|z} + BG \tag{6}$$

where $N_{\text{inc,isotope}}$ and $N_{\text{inc,spin}}$ are the nuclear isotope and spin incoherent cross sections, respectively, $N_{\text{coh}}$ is the coherent nuclear cross section, $M$ is the magnetic cross section, and $BG$ is the background contribution. Supplementary Fig. 6a shows excess scattering in $I^{SF}_{P\|Q}$ above the background level at the two-dimensional AF wave vector at $\omega$ = 48 meV. We attribute this to scattering from magnetic fluctuations. To further confirm this, we measure all six SF and NSF neutron polarization configurations for select wave vectors and extract the pure magnetic intensity from both SF and NSF channels: $M_{P\perp Q} + M_{P\parallel z} = 2\times I^{SF}_{P\parallel Q} - I^{SF}_{P\perp Q} - I^{SF}_{P\parallel z} = I^{NSF}_{P\perp Q} + I^{NSF}_{P\parallel z} - 2\times I^{NSF}_{P\parallel Q}$. Supplementary Fig. 6b shows unambiguous evidence of magnetic scattering at $q_{AF}$ from both SF and NSF scattering

**Supplementary Note 5**

**Effect of superconductivity on the magnetic susceptibility**. HgUD71 does not exhibit a SC resonance. However, we observe subtle changes of the susceptibility ($\Delta\chi''$) in the SC phase at wave vectors away from $q_{AF}$ (Figs. 3b,d and Supplementary Fig. 8). $\Delta\chi''$ around $\omega_1$ is due to a slight increase of momentum width at 5 K (Figs. 1b,c and Fig. 3b inset and Supplementary Fig. 8), whereas at $\omega_2$ it results from an increase in amplitude on the upward dispersive part of the spectrum (Figs. 2c,h,m).

**Supplementary Note 6**

**Temperature dependence of the commensurate response**. Supplementary Fig. 10 shows the evolution of the low-energy magnetic response across the temperatures $T_c$ = 71 K and $T_{CDW}$ = 200 K[7] for HgUD71. In order to reduce background contributions, the scattering at 350 K is first subtracted from that at 5 K, 85 K and 220 K (Supplementary Figs. 10a,b,c). The gapped commensurate $q_{AF}$ response is already observed in the (H,1/2) vs. energy slices. The background subtraction procedure described in Supplementary Note 1 is then applied at all energies to more clearly isolate the AF fluctuations in Supplementary Figs. 10e,f,h. Besides the slightly broader response at 5 K described in the main text, the overall commensurate spectrum remains largely impervious to the onset of superconductivity and CDW order. This is also apparent from the line cuts of the data in Supplementary Figs. 10i. Supplementary Figs. 10d,h shows the enhancement of the response between $T_{CDW}$ and $T_c$. The lack of significant changes in the **Q**-$\omega$ dependence or in the scattering intensity (Fig. 1a) across $T_{CDW}$ and $T_c$ highlights that the commensurate low-energy spectrum is a signature of the PG state which has higher characteristic temperature $T^*$ ($T_c < T_{CDW} < T^*$).

**Supplementary References**